\documentclass[notitlepage,prd,showpacs,superscriptaddress,nofootinbib,floatfix,showkeys,10pt]{revtex4-1}
\usepackage{graphicx}
\usepackage{amsmath}
\usepackage{bm}
\usepackage{yhmath}
\usepackage{mathtools}
\usepackage{wasysym}
\usepackage{subfigure}
\usepackage{xcolor}
\definecolor{CherryRed}{rgb}{.65,0,.2}
\definecolor{RubyRed}{rgb}{.88,0.07,.3}
\definecolor{CralRed}{rgb}{1,0.25,.25}
\definecolor{CobaltBlue}{rgb}{0,0.28,.67}
\definecolor{RoyalBlue}{rgb}{0.25,0.41,.88}
\definecolor{EmeraldGreen}{rgb}{0.31,0.78,.47}
\definecolor{EmeraldGreen}{rgb}{0.31,0.78,.47}
\definecolor{LimeGreen}{rgb}{50,205,50}
\definecolor{ForestGreen}{RGB}{34,139,34}
\definecolor{PineGreen}{RGB}{1,121,111}\usepackage{cases}
\usepackage[colorlinks=true, linkcolor=EmeraldGreen,citecolor=RoyalBlue,urlcolor=ForestGreen,linkcolor=ForestGreen]{hyperref}
\usepackage{subfigure}
\usepackage{times}
\usepackage{booktabs,bm}
\usepackage{slashed}
\usepackage{amsfonts,amssymb,stmaryrd,latexsym,amsmath}
\usepackage{textcomp}
\usepackage{multirow}
\usepackage{cancel}
\usepackage{array}
\usepackage{hyperref}
\usepackage{pgf}
\usepackage{orcidlink}

\def\SU3{{\text{SU(3)}_{\rm F}}}

\def \pcs4338{{P_{\psi s}^\Lambda(4338)^0}}

	\DeclareUnicodeCharacter{200B}{} 
 	 \DeclareUnicodeCharacter{2002}{}

\begin{document}
	
	\title{\textcolor{CobaltBlue}{Total Angular Momentum Assignments in the Hadron Spectrum \\[0.4cm]
    \large{\textcolor{red}{\textbf{The HADLENS\footnote{HADron LEarning Neural Systems} Collaboration:} }}}}

	\author{S.~Rostami\orcidlink{0000-0001-7082-6279}}
	\email{saeedeh.rostami@cern.ch}
	\affiliation{Department of Physics, University of Tehran, North Karegar Avenue, Tehran 14395-547,  Iran }
	\author{ Omid~Zamani~Jamshidi\orcidlink{0009-0004-8944-2784}}
	\email{zamanijamshidi.omid@gmail.com}
	\affiliation{Department of Physics, University of Tehran, North Karegar Avenue, Tehran 14395-547,  Iran }
	\author{B.~Dehghanzad\orcidlink{0009-0005-4096-137X}}
    \email{behrad.dehghanzad@ut.ac.ir}
	\affiliation{Department of Physics, University of Tehran, North Karegar Avenue, Tehran 14395-547,  Iran }
	\author{K.~ Azizi\orcidlink{0000-0003-3741-2167}}
	\email{kazem.azizi@ut.ac.ir}
	\thanks{Corresponding author}
	\affiliation{Department of Physics, University of Tehran, North Karegar Avenue, Tehran 14395-547,  Iran }
	\affiliation{Department of Physics,  Faculty of Engineering and Natural Sciences,  Dogus University, Dudullu-\"{U}mraniye, 34775 Istanbul, T\"urkiye}
	
	\date{\today}

	\begin{abstract}
      Predicting the total angular momentum $J$ for hadronic resonances whose 
spins have not yet been experimentally determined constitutes an 
important step in the spectroscopy of observed hadrons.
In the present study, we develop a data-driven model in order to 
predict the value of $ J $ for hadrons with unknown spins
using a dataset of known mesons and baryons from the PDG,
including information on their quark content, isospin, parity, 
mass and other spectroscopic features. In particular, we employ 
XGBoost classifiers along with nested optimization in order to discover
spectroscopic patterns across different hadronic sectors. We also prepare a systematic list of 
hadrons with unknown or poorly determined values of $ J $, and study spin-parity assignments for those hadrons.
Since the parity hypothesis is taken as an input to the model, 
our model predicts the value of $ J $ conditional on a given parity hypothesis
rather than independently determining the full $J^P$ assignment.
Our predictions for several theoretically studied states show good 
agreement with results from quark models, QCD sum rules, molecular models, 
and other phenomenological approaches, while predictions for some other 
states differ from existing theoretical assignment. These differences motivate 
further theoretical investigations and may be tested by future experimental measurements.
                        	\end{abstract}

	%%%%%%%%%%%%%%%%%%%%%%%%%%%%%%%%%%%%%%%%%%%%%%%%%%%%%%%%%%%%%%%%%%%%%%%%%%%%%%%%%%
	\maketitle
	\renewcommand{\thefootnote}{\#\arabic{footnote}}
	\setcounter{footnote}{0}

	\section{Introduction}
Understanding the hadron spectrum is one of the major challenges of the non-perturbative regime of QCD.
The quark model provides a very successful description of mesons and baryons
\cite{ Gell-Mann:1964ewy,Gell-Mann:1962yej},
while hadron properties are determined by the strong interaction among quarks and gluons.
At the energy scales proper to hadron spectroscopy, non-perturbative QCD effects play 
an important role, making conventional perturbative methods insufficient on their own.
Therefore, conventional perturbation methods are not applicable in this regime \cite{ Peskin:1995ev}.
The spectrum has a rich structure resulting from confinement,
flavor symmetry breaking, spin-dependent interactions and orbital and radial excitations.
Thus, knowing the mass of a hadron resonance is not sufficient to fully characterize it.
Instead,  knowledge of its total angular momentum and parity, $J^P$, 
allows one to identify the hadron in the spectrum, understand its decay properties, and distinguish between several interpretations.
The total angular momentum of a hadron is obtained from the coupling 
of intrinsic spins and relative orbital angular momenta of its constituent particles. 
In the conventional quark model, $S$ and $L$ are coupled such that $J=|L-S|,\ldots,L+S$. 
In a quark-antiquark meson, the parity is given by $P=(-1)^{L+1}$, while in a three-quark baryon, it depends on the orbital angular momenta of its internal quark configurations. 
Therefore,  different orbital excitations can result in different parities, while  different spin couplings could 
give different values of $ J $ even for states with similar masses. The spectroscopic notation $n^{2S+1}L_J$  is commonly used for labelling the configurations of a standard quark model, particularly for conventional mesons. Additionally,  quantum numbers $J^P$ constrain the allowed partial waves  in the strong decays through angular momentum and parity conservation.

Despite significant progress in experimental physics, the quantum numbers of many hadronic states are still not fully known.
The Particle Data Group (PDG) lists a large number  of meson and baryon resonances with measured mass but unknown or poorly constrained total angular momentum $ J $,  parity or both  \cite{ParticleDataGroup:2024}.
This uncertainty is present across different sectors of the hadron spectrum.
For example, several excited mesons and baryons in the light and 
strange sectors still have multiple possibilities for their spin-parity quantum numbers.
Similarly, in the charm and bottom sectors the spectrum has been 
significantly extended in the last two decades and the quantum numbers of several  remain unclear.
The problem gets even more challenging for exotic candidates like hidden-charm 
pentaquarks and other multiquark states. Here, different internal structures 
can lead to  different quantum number assignments.
For instance, according to the hadronic molecular approach, the total 
angular momentum arises  from combining the spins of the constituent 
hadrons with their orbital angular momentum,  whereas the parity 
is given by $P=P_1P_2(-1)^L$. Therefore, the same hadronic constituents 
may allow different $J^P$ values  depending on their spin coupling and
orbital configurations. This is one reason why different theoretical pictures can 
lead to different interpretations of the same experimentally observed resonance.

A number of complementary theoretical approaches have been applied to 
the spectroscopy and quantum numbers of hadronic states. 
For instance, constituent quark models classify conventional mesons 
and baryons based on their radial and orbital excitations and the 
coupling of the constituent quarks spins. Moreover, these approaches 
can be used to investigate the $J^P$ values using mass spectra and strong decay patterns \cite{Klempt:2009pi,Vijande:2004he,Godfrey:2019cmi,Wang:2017kfr}.  
QCD sum rules is also one of the methods which have been used to describe hadron spectroscopy. According to this method, interpolating currents with certain quantum numbers are used to construct correlation functions which allow studying the properties of the corresponding states. Such an approach has also been applied for conventional heavy baryons as well as for exotic tetraquarks and pentaquarks \cite{Azizi:2020azq,Aliev:2018vye,Wang:2019tlw,Wang:2019got,Ben:2025vfw,Agaev:2022ast,Agaev:2023wua}.
The molecular and coupled-channel approaches become especially important for hadronic states which are near the hadronic thresholds. In such approaches, the states can appear dynamically due to interactions between hadronic channels \cite{Ramos:2002xh,Sekihara:2015qqa,Feijoo:2023wua,Xiao:2019mvs,Xiao:2021rgp}.
These distinctions between the models can result in different assignments for the same resonance. 
For instance, $\Xi_b(6227)$ has been studied as a conventional $P$-wave bottom baryon with possible assignment of $J^P=3/2^-$ or $5/2^-$ \cite{Wang:2018fjm,Cui:2019dzj}. 
Molecular interpretations  have also been considered for the $\Xi_b(6227)$ \cite{Huang:2018bed},
 which is an example of how the assumptions on the underlying structure influence its quantum numbers. 
 Different excitation mechanisms have been assumed for $B_J(5970)$
  leading to possible assignments of $1^-$, $2^-$, 
  and $3^-$ \cite{Xu:2014mka,Chen:2021qkx,li:2021hss}. 
  These cases serve as an example for how theoretical predictions 
  should not be considered  definitive classifications but rather as 
  model dependent interpretations that need to be tested against experimental data.

Theoretical spin assignments for a resonance should always 
be considered as hypotheses until they are experimentally established. 
Predictions of the value of $J$ under a given parity value can provide 
an additional aspect for comparison with theoretical calculations.
Agreement between data-driven predictions  and theoretical calculations 
is meaningful, and discrepancies are worth attention.

Machine learning (ML) algorithms have a long history in high-energy physics \cite{Carleo:2019ptp}. 
Initially, they were mostly used as multivariate algorithms for selecting events, 
separating signal from background and classifying particles \cite{TMVA:2007ngy}.
With the emergence of deep learning, new applications such as classification of jets, 
event reconstruction and detector simulation have appeared \cite{Guest:2018yhq,Radovic:2018dip}. 
Also, Generative adversarial networks (GANs) have been applied to generate jet images and simulate particle showers in calorimeters \cite{deOliveira:2017pjk,Paganini:2017hrr,Paganini:2017dwg}.

Lately, ML methods have even found use in hadron spectroscopy, 
in which case ML methods can be used to detect any correlations 
that exist between hadron spectra and investigate hadronic states whose 
properties have yet to be elucidated \cite{Liu:2022uex,Ng:2021ibr}. 
In fact, these methods have found applications in investigating near threshold 
resonances, line shapes and poles, as well as possible spin values for some exotic hadrons \cite{Ng:2021ibr,Zhang:2023czx,Co:2024bfl,Frohnert:2025usi}. 
Other uses of ML in hadron spectroscopy have included prediction of the 
width of mesons and investigations of multiquark systems and QCD \cite{Tong:2025jnd,Wu:2025wvv,Aarts:2025gyp,ThomasArun:2025uyi}.

In our previous works, we  used  deep neural networks (DNN) 
and conditional generative adversarial networks (cGANs)
 to predict meson masses and decay widths, as well as the study of fully-heavy tetraquarks. 
 Later, similar investigations were pursued on heavy baryons and pentaquarks using DNN models and Transformer \cite{Malekhosseini:2024eot,Rostami:2025sff,Malekhosseini:2025hyx,Rostami:2026jyz}.

In the present study, we extend our previous works by predicting
the total angular momentum $J$ of hadron resonances whose
 spin assignments  remain unknown or poorly constrained. 
 For this purpose, a novel XGBoost framework utilizing nested optimization, 
  and graph-assisted modelling has been developed. The ML 
 model is trained with the help of the available list of mesons and baryons 
 reported by PDG along with their quark content, isospin, parity, and other spectroscopic properties.
 The above model is then used to make predictions regarding the spin assignments of 
  hadronic states with unknown spins, and those predictions are compared with current interpretations. 
 Since parity is used as an input, the model predicts $J$ under a given parity 
 assumption rather than independently determining the full $J^P$ assignment. 
 The prediction of the model for the states whose parities are not known can be 
 made using different assumptions of parity.
 
This paper is organized follows.
 In Section \ref{Method}, we introduce an XGBoost-based machine learning 
 framework that incorporates genetic search and nested cross validation. In Section \ref{Res}, 
 we present predictions for the spins of mesons, baryons, and candidates for exotic 
 hadrons, comparing them with theoretical assignments found in the scientific literature. 
 Finally, in Section \ref{SC}, we conclude this paper by summarizing our results, discussing 
the physical implications of our findings, and outlining potential avenues for future investigation.

\section{machine learning framework: XGBoost}\label{Method}
In this study, we use a supervised machine learning
 approach to predict the total angular momentum quantum number
$J$ of hadrons. As $J$ takes only integer or half-integer
values, the prediction problem is a multiclass classification
problem rather than a regression problem. The classifier was trained using
meson and baryon states with experimentally established $J$ values 
and was then applied to predict the $J$ values of hadronic resonances having
unknown spin values. Our method is based on the XGBoost
classifier trained via hyperparameter optimization  and cross validation.

Each hadronic state has been described using its constituent-quark content and a compact set of experimentally reported spectroscopic quantities. The initial input vector is
\begin{equation}
\label{eq:inputs}
\mathbf{x}=(d,\bar d,u,\bar u,s,\bar s,c,\bar c,b,\bar b,n_{valence},I,P,M).
\end{equation}

Here the first ten components are quark content,  $n_{valence}$ is the total number of valence constituents,
 $I$ is isospin, $P$ is parity, and $M$ is the measured hadron mass. As mentioned above, parity is provided as an input hypothesis rather than predicted by the model.
Integer and half-integer spins are represented by the integer class label:
\begin{equation}
 y=2J.
\end{equation}
The training sample includes 16 spin classes,
\begin{equation}
J=0,\frac{1}{2}, 1, \frac{3}{2}, 2, \frac{5}{2}, 3, \frac{7}{2}, 4, \frac{9}{2}, 5, \frac{11}{2}, 6, \frac{13}{2},  7, \frac{15}{2}.
\end{equation}
This encoding represents both integer and half-integer spin values using integer class labels.

XGBoost is a machine learning algorithm that
employs an ensemble of decision trees \cite{Chen:2016btl}.
This algorithm falls under the category of gradient boosting algorithms,
where a number of weak learners are used to build a more precise
model. In contrast to the approaches where all decision trees
are trained separately, XGBoost constructs trees sequentially. At each step
a new tree is grown in order to enhance the prediction of the current
ensemble by minimizing an objective function.

A decision tree partitions the input feature space
into different regions using the conditions applied to the input variables.
Each leaf of the tree contributes to the prediction of the model.
By combining multiple trees, XGBoost can capture complex and non-linear dependencies
between input features and target variables. This algorithm includes
regularization terms in order to control the complexity of the model
and reduce the risk of  overfitting. These features make XGBoost suitable for
datasets containing both discrete and continuous variables.

XGBoost can  be used for regression as well as
classification problems. In the case of multiclass
classification, the model computes a score for each
possible class and converts the scores into class
probabilities through a softmax function. The
predicted class is then selected based on the resulting probability distribution.

\begin{figure}[t]

  \centering

  \includegraphics[width=0.965\linewidth]{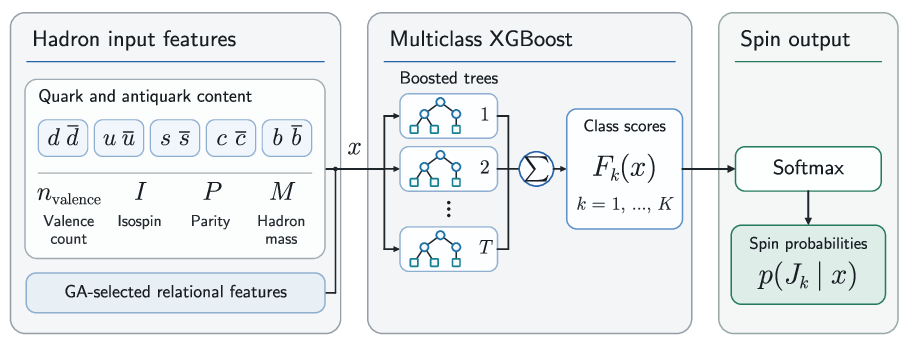}

  \caption{Schematic representation of a multiclass XGBoost classifier used for spin prediction.}
  \label{fig:xgboost}
\end{figure}

The XGBoost model combines the outputs of multiple
regression trees in an additive manner
\cite{Chen:2016btl}. The general tree ensemble formulation is
expressed as:
\begin{equation}
\hat{y}_i = \phi(\bf{x}_i)
= \sum_{k=1}^{K} f_k(\bf{x}_i),
\qquad f_k \in \cal{F},
\end{equation}
where $K$ is the number of trees, $f_k$ denotes the function corresponding to the $k$-th tree,
and $\mathcal{F}$ represents the space of regression trees. The structure of the XGBoost used for spin prediction is illustrated in Fig.~\ref{fig:xgboost}.

The XGBoost classifier is trained by minimizing a regularized objective function, which combines a loss term with a penalty for model complexity \cite{Chen:2016btl}. In multi-class classification, logarithmic loss (log loss) is used to evaluate the predicted probabilities for the classes; consequently, a heavier penalty is imposed if the model assigns a low probability to the correct class.
The probabilities predicted in this study are useful, as the spin assignment for certain hadronic states is subject to uncertainty. For these states, a probability distribution over possible values ​​of $ J $ provides more information regarding the model's predictions than the assignment of a single class.

The XGBoost model is well-suited for the current dataset, which comprises discrete quark counts, quantum numbers, and continuous mass values. This tree-based model can extract non-linear relationships among these variables without requiring an explicit analytical relationship between the input features and $J$. 

For the XGBoost stage, the preprocessing retains the original variables in their physical scales; tree-based classifiers do not require global standardization. Missing numerical values ​​are passed directly to XGBoost, which learns a default routing strategy for missing inputs within the individual trees. This avoids the need to assign an arbitrary numerical substitute for an unmeasured or unreported physical quantity.

The classifier's performance depends on the complexity of trees, the learning rate, sampling fractions, and the level of regularization. Optuna automatically tunes all of these parameters~\cite{Akiba:2019lwq}. In each stage of training, the Optuna study minimizes the average multi-class log-loss in the inner cross-validation folds. A Tree-structured Parzen Estimator (TPE) sampler searches through the parameter space, and some poorly performing trials may be cut off from the search.

The search ranges for each parameter are listed in  Table~\ref{tab:hyperparameters}. The learning rate $\eta$, minimum child weight, and regularization coefficients are drawn from log-scales when appropriate. Instead of specifying the number of boosting rounds upfront, the algorithm tunes it along with other tree parameters. Finally, the tuning process runs 20 trials of Optuna for each outer fold.

\begin{table}[t]
  \centering
  \begin{tabular}{lll}
   \hline
    Parameter & Search range & Role \\
 \hline
    $\eta$ & 0.01-0.18 (log scale) & learning rate \\
    \texttt{max\_depth} & 2-6 & maximum tree depth \\
    \texttt{min\_child\_weight} & 0.5-20 (log scale) & minimum child weight \\
    \texttt{subsample} & 0.65-1.00 & row subsampling \\
    \texttt{colsample\_bytree} & 0.60-1.00 & column subsampling \\
    $\gamma$ & 0-5 & split regularization \\
    $\alpha$ & $10^{-8}$-10 (log scale) & $L_1$ regularization \\
    $\lambda$ & $10^{-3}$-100 (log scale) & $L_2$ regularization \\
    \texttt{max\_delta\_step} & 0-5 & update constraint \\
    boosting rounds & 80-1000 (log scale) & ensemble size \\
    \hline
  \end{tabular}
  \caption{XGBoost hyperparameters optimized with Optuna.}
    \label{tab:hyperparameters}
\end{table}

\subsection{Model training and validation}

A genetic algorithm provides an additional stage of model selection. All 14 initial hadronic variables in Equation \eqref{eq:inputs} are retained throughout the process while genetic algorithm selects from an additional set of features. The genetic chromosome controls only the selectable portion of the input set. By default, a population consisting of 16 candidate configurations evolves over seven generations. In each generation, the top three candidates are preserved as elites, and tournament selection is performed using groups of three candidates. The crossover and mutation probabilities are set to 0.80 and 0.035, respectively. Following this screening phase, the three leading candidates are each refined through six Optuna trials, after which a final optimization of the XGBoost model is conducted.

The genetic fitness function places the greatest emphasis on classification accuracy. Moreover, it includes probability quality as well as performance in regions of the known data that are most similar to the target states without using the unknown spin values of the target states.  The principal accuracy criterion is a combination of overall accuracy, accuracy weighted by similarity to the target domain, and stress zone accuracy with respective weights of 0.30, 0.30, and 0.40. Some fine-tuning according to the negative log-likelihood, Top-2 accuracy, and the size of the input for the model is applied as a tiebreaker. The goal is to select configurations that perform well within the target domain, without utilizing the unknown spin values ​​of the target states.

The comprehensive evaluation employs nested cross-validation using 343 physical resonances. First, the physical resonances are split into five outer folds. Then, for each outer fold, the rest of the data is split into four inner folds for genetic selection and Optuna optimization of hyperparameters. It is done on the level of physical resonance, so that all rows corresponding to one physical resonance fall into one fold. 
This ensures that various inputs corresponding to a same physical resonance do not appear simultaneously in both the training and validation sets.

Nested tuning is especially suitable for such a small spectroscopy dataset. All choices of hyperparameters are done within the training folds, and the labels of the outer validation folds are not used in the model selection step.

\begin{figure}[h!]

  \centering

  \includegraphics[width=0.6\linewidth]{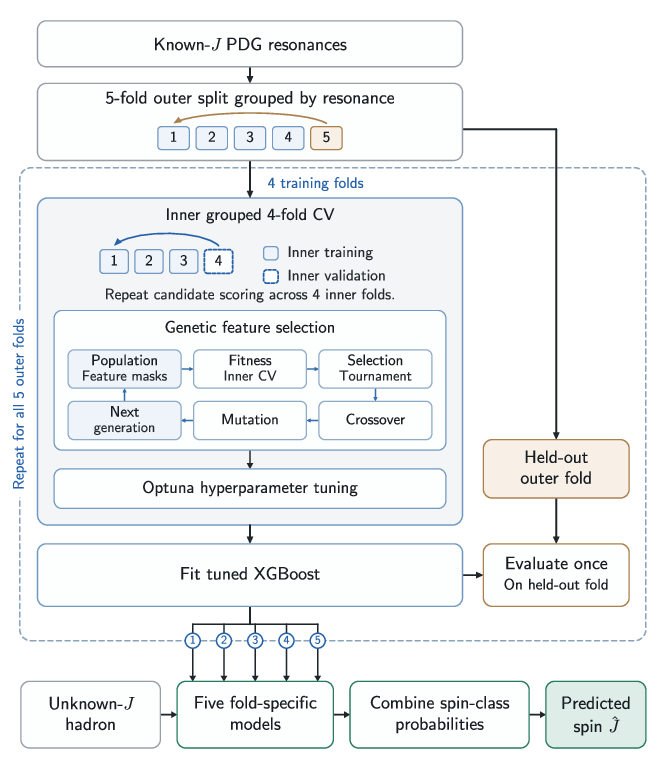}

  \caption{The nested training and optimization workflow.}
  \label{fig:nested}
\end{figure}

In case there are several rows for a particular physical resonance, each row is allocated a training weight of the inverse of the number of rows representing the physical resonance. Therefore, all physical resonances, irrespective of the number of occurrences in the source table, contribute equally to the overall statistical weight. The nested training and validation procedure is illustrated in Fig.~\ref{fig:nested}.

The five fold-specific models from outer cross-validation form the deployment ensemble. Their probability outputs are combined at the time of prediction. This reduces dependence on any single training partition of moderate size and provides more stable spin assignments for unknown states.

The parity $P$ is provided as an input condition, rather than being inferred by the classifier. Consequently, the output should be interpreted as $P(J\mid\mathbf{x})$, where $ \mathbf{x} $ includes the assumed parity, rather than as a simultaneous determination of $J^P$. In this implementation, if the parity of a target state is unavailable, hypotheses corresponding to positive and negative parity are evaluated separately. Both outputs are retained as conditional predictions; however, neither provides a probability for the parity itself.

\section{Results}\label{Res}

In this section, we describe the spin-parity prediction obtained using the machine learning model for the hadronic states whose spin-parity quantum numbers have not yet been firmly determined experimentally. Here, we consider both conventional meson and baryon states, as well as several candidates of exotic hadrons. As many of them have not yet been assigned experimentally, the prediction of the model is compared primarily with the results that exist in the theoretical literature.

Regarding spin parity, the comparison is not as straightforward as with mass, where the difference between predicted and reference values can be estimated numerically. Spin-parity quantum numbers are discrete, and for many poorly established states, their theoretical assignments depend strongly on the approach used. In this way, we compare our results with the predictions from various models such as quark models, QCD sum rules, molecular models, effective field theory calculations,  and many more. Agreement among several independent theoretical approaches provides a useful benchmark for assessing our prediction. However, sometimes, different approaches give different $J^P$ values, and this should be taken into account.

In the current approach, parity is used as an input assumption, and the model predicts the associated spin value. Thus, the predictions obtained for the positive and negative parity branches should not be interpreted as probabilities for the parity of a given state. In case there is good motivation for a certain parity choice based on theoretical or experimental data, the corresponding parity branch is selected, and the calculated spin is compared with the literature. If both parities are still reasonable in theoretical terms, both predictions by the model are kept, and no unique $J^P$ assignment is selected.

Thus, the following degrees of consistency between our results and the literature are used below. Exact agreement with a theoretical assignment refers to the cases when the predicted spin and parity agree with some specific theoretical hypothesis, while partial consistency occurs when parity agrees with theory but the spin does not.
For states with several theoretical interpretations, agreement with one of them is regarded as compatibility with that scenario rather than as a unique identification of the state. Finally, when there are not enough constraints on parity and/or spin from either theoretical or experimental studies, the prediction is presented without any preferred $ J^P $ assignment. The following subsections discuss these patterns separately for mesons and exotic candidates, light strange baryons, and heavy baryons and hidden charm pentaquarks.

\begin{table}[h!]

\centering

\begin{tabular}{l c c}

\hline

State & Our model & Theory \\

\hline

$K(1630)$ & $1^-$ & $1^-$ \cite{Vijande:2004he} \\
\\ \hline
$B_J(5970)$ & $1^-$ & $1^-$ \cite{Xu:2014mka,Godfrey:2019cmi}, $2^-$ \cite{Chen:2021qkx},
$3^-$ \cite{li:2021hss} \\ \\ \hline
$B_{sJ}(6063)$ & $1^-$ & $1^-$ \cite{Garg:2023hvp,Gandhi:2022nnk}, $3^-$ \cite{Hao:2022ibj} \\ \\ \hline
$B_{sJ}(6114)$ & $1^-$ & $1^-$ \cite{Garg:2023hvp,Gandhi:2022nnk,Hao:2022ibj,Zhang:2026duk},
$2^-$, $3^+$ \cite{Zhang:2026duk} \\ \\ \hline
$T_{c\bar{c}}(4020)$ & $1^+$ & $1^+$ \cite{Wang:2013exa,Wang:2022clw,Wua:2023ntn},
$0^+,\,2^+$ \cite{Wang:2014gwa}\\ \\ \hline
$T_{c\bar{c}}(4050)$ & $1^+$ & $1^+,\,2^+$ \cite{Wang:2019tlw} \\ \\ \hline
$T_{c\bar{c}}(4055)$ & $1^+$ & $1^+$ \cite{Wang:2019tlw} \\ \\ \hline
$T_{c\bar{c}}(4100)$ & $1^+$ & $0^+,\,1^+$ \cite{Wang:2019mta} \\\\  \hline
$T_{c\bar{c}}(4250)$ & $1^\pm$ & $1^+,\,1^-$ \cite{Cao:2018vmv} \\ \\ \hline
$T_{b\bar{s}}(5568)$ & $0^+$ & $0^+,\,1^+,\,2^+$ \cite{Wang:2016mee,Stancu:2016sfd} \\ \\ \hline
$D^{*}(2640)$ & $1^-$ & $0^-$ \cite{Vijande:2004he}, $1^-$ \cite{DELPHI:1998oyl}, $2^-,\,3^-$ \cite{Melikhov:1998jm,Page:1998pp} \\ \\ \hline
$D_{sJ}(3040)$& $1^\pm$ & $1^+$ \cite{Sun:2009tg,Jiang:2024ftx}, $2^-$ \cite{Colangelo:2010te} \\  \\ \hline
$B_J(5840)$& $0^-$& $1^-$ \cite{Gupta:2018xds,Yu:2019iwm}, $0^-$ \cite{li:2021hss} \\  \\ \hline
$f_J(2220)$& $2^+$& $2^{++}$ \cite{Hsiao:2014tda} \\  \\ \hline
%$X(4350)$& $2^+$& $0^+,\,2^+$ \cite{Belle:2009rkh,Ma:2010xx};$2^{++}$ \cite{Liu:2009fe} \\  \\ \hline
\hline

\end{tabular}

\caption{Comparison of the model predictions with representative theoretical
assignments for mesons and exotic-hadron candidates.}

\label{tab:meson_comparison}

\end{table}

Table \ref{tab:meson_comparison},  shows comparisons of the model predictions with existing  theoretical assignments for the mesons and exotic-hadron candidates. The spin-parity quantum numbers of the $ K(1630) $ state have not been
experimentally established. The current PDG listing gives $ I(J^P) = 1/2(?^?) $,
 with a mass of $ 1629 \pm 7 $ MeV. Therefore, any $ J^P $ assignment used for $ K(1630) $ should be regarded as a theoretical
interpretation rather than an experimentally established value.
In Ref.~\cite{Vijande:2004he} the authors studied the
light-strange meson spectrum within a constituent quark model. They
identified $ K(1630) $ with the $ 2^3S_1 $ state and predicted
$ J^P = 1^- $. Their calculated mass, 1620 MeV, is close to the
experimental value. 
In Ref.~\cite{Taboada-Nieto:2022igy}, the author analyzed the kaon spectrum
using a constituent quark model. Their analysis does not support a
straightforward conventional $ q \bar{q} $ assignment of $ K(1630) $ as a
pseudoscalar state. In particular, in their discussion of the excited
kaon spectrum, the authors point out the ambiguity in the
interpretation of $ K(1630) $ and that its quantum numbers need to be examined
further.
In a different interpretation, $ K(1630) $ has also been considered as
a candidate for a strange pseudoscalar state in the second radial
excitation nonet. This interpretation corresponds to $ J^P = 0^- $,
but it should be regarded as a candidate assignment rather than a
well-established quantum number.
Thus, the theoretical literature cannot give a single interpretation of the
quantum numbers of $ K(1630) $. As a result, our model prediction, $ J = 1$
for the negative parity. This result  agrees with the assignment of Ref.~\cite{Vijande:2004he}.
Although, this agreement dose not  establish a unique identification of the state.

Several theoretical studies suggest various spectroscopic
assignments of  the $ B_J(5970) $ state.
According to Ref. \cite{Xu:2014mka}, Xu, Liu and Matsuki  interpreted the $ B_J(5970) $ state as the 
$ 2^3 S_1 $ state of the B meson family, which
corresponds to $ J^P = 1^- $.
Godfrey and Moats \cite{Godfrey:2019cmi}, based on the relativized quark model,
have considered the $ B_J(5970) $ most likely to be the $ B^*[2^3S_1] $ state,
corresponding to $ J^P = 1^- $.
On the other hand, Ref. \cite{li:2021hss} proposed a different assignment,
namely that $ B_J(5970) $ is the $ 1^3D_3 $ state, which corresponds
$ J^P = 3^- $. However, they pointed out that the pure $ 2^3S_1 $ state assignment
$ J^P = 1^- $ could not be excluded.
Based on the chiral effective Lagrangian approach, it has been determined in Ref. \cite{Chen:2021qkx} that the quantum numbers of the $ B_J(5970) $ state are $ J^P = 1^- $ or $ 2^- $.
Therefore, the existing theoretical literature does not offer any
unique $ J^P $ quantum number for $ B_J(5970) $. The $ J^P = 1^- $ quantum number assignment is based on several independent
studies, while the $ J^P = 3^- $ quantum number assignment
is another option. In our analysis, under the assumption of negative parity, we predict a value of $J=1$, corresponding to $J^P=1^-$. This result is consistent with the assignments proposed by Xu et al. \cite{Xu:2014mka,Godfrey:2019cmi}. However, this agreement does not rule out alternative attributions that have been proposed in scholarly texts.

The recently observed $B_{sJ}(6063)$ and $B_{sJ}(6114)$ states have been 
investigated using various theoretical approaches. According to HQET, 
these two states are considered to be $2^3S_1$ and $1^3D_1$ 
excitations, that correspond to $J^P=1^-$ \cite{Garg:2023hvp,Gandhi:2022nnk}. 
However, other calculations for $B_{sJ}(6063)$ suggest a $1^3D_3$ assignment with $J^P=3^-$,
while assigning $J^P=1^-$ to $B_{sJ}(6114)$ \cite{Hao:2022ibj}. 
In another study, where $B_{sJ}(6114)$  considered to be a molecular state, also supports the $J^P=1^-$ 
assignment based on its decay width, although  $J^P=2^-$ and $J^P=3^+$ remain possible \cite{Zhang:2026duk}. Under the negative-parity hypothesis, our model predicts $J=1$ for both states $B_{sJ}(6063)$ and $B_{sJ}(6114)$,
 consistent with the $1^-$ assignments proposed in these studies.
 
For the $T_{c\bar{c}}(4020)$ states, our model predicts $J^P=1^+$ for 
all three charge states, $T_{c\bar{c}}(4020)^{+}$, $T_{c\bar{c}}(4020)^{0}$, 
and $T_{c\bar{c}}(4020)^{-}$. This result is consistent with several theoretical analyses  
of the corresponding $Z_c(4020)$ state.In particular,  QCD sum rule analyses 
in both molecular and tetraquark frameworks have favored the axial vector 
assignment $J^{PC}=1^{+-}$ \cite{Wang:2013exa}, while molecular 
calculations have also predicted $J^{PC}=0^{++}$ and $2^{++}$ configurations \cite{Wang:2014gwa}.
Moreover, the value of $J=1$ is confirmed  by a constituent quark model 
with instanton induced interactions, which finds a $1^{+-}$ state near 
mass of $Z_c(4020)$  \cite{Wang:2022clw}, and by QCD sum 
rules identifying $1^{+-}$ states around $4.01$--$4.02$ GeV 
as possible $Z_c(4020/4055)$ candidates \cite{Wang:2019tlw}. 
Recently, an NLO QCD sum rule analysis with operator mixing
 obtained $4.03^{+0.06}_{-0.07}$ GeV and assigned $J^P=1^+$ to 
 the $Z_c(4020)$ \cite{Wua:2023ntn}.
In general, our prediction $J=1$ is in good agreement with the  
interpretation of the $Z_c(4020)$ from theoretical point of view.  
Though some QCD sum rule approaches suggest $J=2$, the 
agreement with several independent methods provides 
support for the $J=1$ quantum number obtained within our data-driven model.

Moreover,
for the higher mass $T_{c\bar{c}}$ states, our model predicts $J=1$ for $T_{c\bar{c}}(4050)^{+}$, $T_{c\bar{c}}(4055)^{+}$, $T_{c\bar{c}}(4100)^{+}$, and $T_{c\bar{c}}(4250)^{+}$. According to theoretical studies, different assignments have been proposed for these states. In particular, QCD sum rules predict $J^{PC}=1^{++}$ and $2^{++}$ configurations for $Z_c(4050)$, while a $1^{+-}$  has been obtained for $Z_c(4055)$ \cite{Wang:2019tlw}. As for $Z_c(4100)$, the QCD sum rules have considered scalar and vector tetraquark interpretations, with a mixed scalar configuration providing a possible description near 4.1 GeV \cite{Wang:2019mta}. Finally for $ Z_2(4250) $, a spin-parity analysis allows  $J^P=1^+$ or $1^-$ \cite{Cao:2018vmv}, in agreement with our prediction of $J=1$.

Our model also predicts $J^p=0^+$ for $T_{b\bar{s}}(5568)^{+}$, commonly associated with the proposed $X(5568)$ state. Theoretical studies have considered several possible assignments for this state, including $J^P=0^+$ and $1^+$, while a constituent quark model has predicted  $0^+$, $1^+$, and $2^+$ tetraquark configurations \cite{Wang:2016mee,Stancu:2016sfd}. Thus, our $J^p=0^+$ prediction is consistent with the theoretical possibilities, although the spin assignment of $X(5568)$ is model-dependent.
Generally, these comparisons demonstrate that the $J=1$ predictions of our model are consistent with several theoretical possibilities, particularly for $Z_c(4055)$ and $Z_2(4250)$. Regarding $Z_c(4050)$, $Z_c(4100)$, and $X(5568)$, however, alternative values of $J$ have also been suggested, indicating that these states remain theoretically less constrained.

The $D^{*}(2640)$ state provides another example for which a different spin-parity assignment has been proposed.
In 2005 the  Ref.~\cite{Vijande:2004he} studied the state within a constituent quark model and  associated with a radially excited $2S$ pseudoscalar configuration and assigned $J^P=0^-$.
Nonetheless, this is not the only possible interpretation of the state. The DELPHI Collaboration reported evidence for a narrow structure at a mass of $2637\pm2\text{ (stat.)}\pm6\text{ (syst.)}$ MeV, interpreting it as compatible with a radially excited vector $D^{*\prime}$ state with $J^P=1^-$ \cite{DELPHI:1998oyl}. Subsequent theoretical studies questioned this assignment. Ref.~ \cite{Melikhov:1998jm} argued that the very small reported width is difficult to reconcile with a simple radial excitation and discussed alternative interpretations involving orbitally excited states with $J^P=2^-$ or $3^-$. Page \cite{Page:1998pp}, using the framework of heavy quark symmetry, proposed multiple possible assignments, namely, a radial excitation and higher spin orbital excitations. The OPAL Collaboration subsequently reported no evidence for a narrow radial excitation in its search near the predicted mass of $2629$ MeV \cite{OPAL:2001xfs}.  Our model predicts $J^P=1^-$
which reproduces one of the theoretically proposed assignments for this state.

In the case of $D_{sJ}(3040)^+$, the decay pattern seems to suggest an unnatural parity, and both $1^+$ and $2^-$ assignments have been considered \cite{BaBar:2009rro,Colangelo:2010te}. However, several subsequent studies support interpreting the state as a radially excited or mixed $2P$ axial vector state with $J^P=1^+$ \cite{Sun:2009tg,Jiang:2024ftx}. As for positive parity, our model gives $J^P=1^+$, which is in agreement with the theoretical results, while for negative parity it predicts $J^P=1^-$. Therefore, the positive parity  branch reproduces the preferred $1^+$ assignment exactly,
whereas our negative parity prediction, $1^-$, is inconsistent with existing theoretical results.

For $B_J(5840)$, several theoretical studies prefer a negative parity $J^P=1^-$ assignment, although its internal configuration remains uncertain. An HQET analysis prefers a $1D,1^-$ interpretation \cite{Gupta:2018xds}, while a $^{3}P_{0}$ strong decay analysis concludes that it is a radially excited $2^{3}S_{1}$ state, also with $J^P=1^-$ \cite{Yu:2019iwm}. A later quark model analysis likewise allows a pure $2^{3}S_{1}$ state or a $2^{3}S_{1}$--$1^{3}D_{1}$ mixed state with $J^P=0^-$\cite{li:2021hss}. For negative parity, our model predicts $J=0$, corresponding to  $J^P=0^-$.
Our prediction is consistent with the  $2^1S_0$ interpretation considered  in Ref.~\cite{li:2021hss}. Therefore experimental information, specially on the decay channels, is needed to distinguish between these results.

Regarding $f_J(2220)$, the quantum numbers and even the experimental status of the state are not clear yet. It has frequently been discussed as a tensor glueball candidate with $J^{PC}=2^{++}$, motivated in part by lattice QCD expectations for a tensor glueball near this mass region. For instance, Ref.~\cite{Hsiao:2014tda} explicitly treats $f_J(2220)$ as a $2^{++}$ tensor glueball. Our model predicts $ J^P=2^+ $, which is consistent with the available theoretical interpretation of this state.
As well, several experimental searches, including CLEO and BaBar, did not confirm a significant $f_J(2220)$ signal \cite{CLEO:2002uux,BaBar:2007esv}. The result is therefore regarded as an agreement with the commonly discussed $2^{++}$  rather than evidence against the tensor glueball scenario.

\subsection{Light strange baryons}

The model predictions compared with  existing theoretical
assignments for the light strange baryons is depicted in Table \ref{tab:strange_baryon_comparison}.
The $\Xi(1620)$ and $\Xi(1690)$ provide two clear agreements with theoretical descriptions of light strange baryons. The authors of Ref.~ \cite{Ramos:2002xh} dynamically generated the $\Xi(1620)$ as an $s$-wave resonance with $J^P=1/2^-$. More recent chiral unitary calculations also support a dynamically generated molecular interpretation of both $\Xi(1620)$ and $\Xi(1690)$ \cite{Feijoo:2023wua}. For negative parity, our model predicts $ J^P=\frac{1}{2}^{-} $
in exact accordance with the above theoretical assignment.

Regarding $\Xi(1690)$, Sekihara \cite{Sekihara:2015qqa} interpreted the state as an $s$-wave $\bar K\Sigma$ molecule with $J^P=1/2^-$, an assignment consistent with the observed $S$-wave contribution in BaBar data \cite{BaBar:2008myc}. Our model also predicts $ J^P=\frac{1}{2}^{-} $
in agreement with both the theoretical description and the available experimental indication.

There are several possible spin-parity assignments
for $\Xi(2120)$. In the study based on a molecular $\bar K^{*}\Sigma$ structure, the possible quantum numbers were $J^P=1/2^-$, $3/2^-$, and $3/2^+$ as possible states \cite{Huang:2018uox}, whereas an independent coupled channel analysis also associated a dynamically suggested $J=3/2$ state associated with $\Xi(2120)$ \cite{Khemchandani:2017wfb}. Recently, molecular calculations  confirmed $J^P=3/2^-$ as a plausible interpretation \cite{Hei:2023eqz}. Our model predicts $J=3/2$ for both parity hypotheses, corresponding to $J^P=3/2^+$ or $3/2^-$. Thus, the predicted spin agrees with several theoretical interpretations, but the parity is undetermined theoretically.

The broad structure denoted by $\Xi(1950)$ can actually include more than one resonance. Ref.~\cite{PavonValderrama:2011gp} suggested three possible states in this mass region with $J^P=1/2^-$, $5/2^+$, and $5/2^-$. A chiral quark model analysis also found a $J^P=1/2^-$ state to be a possible component of $\Xi(1950)$ \cite{Xiao:2013xi}, while a recent pentaquark study interpreted it as a $\Lambda\bar K^{*}$ state with $J^P=1/2^-$ \cite{Yan:2024usf}. Our model predicts $J=1/2$ for both positive and negative parity. Selecting the negative parity branch therefore gives $J^P=\frac{1}{2}^{-}  $
in agreement with the theoretical interpretations. However, since the experimental quantum numbers remain undetermined and more than one resonance might  contribute in this mass region, the above assignment  cannot be considered unique.

Some recent QCD sum rule calculations provide useful benchmarks for determining the quantum numbers of some of the strange baryons which have not been established yet. In Ref.~ \cite{Ben:2025vfw} propose  the quantum number of $J^P=3/2^-$ for $\Sigma(2250)$, $J^P=1/2^-$ for $\Sigma(2455)$, $\Sigma(2620)$, and $\Xi(2370)$, and $J^P=3/2^+$ for  $\Xi(2250)$. For these states, our model predicts:
\begin{equation*}
J^P(\Sigma(2250)) = \frac{5}{2}^-,\qquad
J^P(\Sigma(2455)) = \frac{1}{2}^-, \qquad
J^P(\Sigma(2620)) = \frac{1}{2}^-
\end{equation*}
\begin{equation*}
J^P(\Xi(2250)) = \frac{3}{2}^{+}, \qquad
J^P(\Xi(2370)) = \frac{3}{2}^{-}, 
\end{equation*}
Our predictions for three of these states align with the assignments proposed in Ref.~\cite{Ben:2025vfw}. However, a discrepancy arises regarding $\Sigma(2250)$ and $\Xi(2370)$: our model predicts $J^P=5/2^-$ and $J^P=3/2^-$, respectively instead of $3/2^-$  and $1/2^-$,assignment proposed in that study.

 The quantum numbers of $\Xi(2500)$ remain poorly established. According to our model, $J = 3/2$ for both $\Xi(2500)^0$ and $\Xi(2500)^-$ under the conditions of positive and negative parity. 
 Since no reliable theoretical assignments for the specific cases under comparison could be found, these predictions remain subjects for future theoretical and experimental investigation.

As for $\Sigma(3000)$, a recent investigation of the $K^-p\rightarrow\phi\Sigma^0$ reaction examined the assignments $J^P=1/2^\pm$, $3/2^\pm$, and $5/2^\pm$, and came to the conclusion that $J^P=5/2^-$ provides a reasonable overall description of the available data \cite{Kim:2026ejb}. However, the authors emphasized that the present measurements are insufficient for a definitive determination. Using the negative parity branch, our model predicts $ J^P(\Sigma(3000))=\frac{1}{2}^{-} $.
In other words, the predicted parity agrees with the most probable  hypothesis, whereas the spin differs from the proposed $5/2^-$ assignment.
Regarding $\Sigma(3170)$,
the quantum numbers of the spin-parity remain  experimentally undetermined and are still presented in the list of the PDG without 
the determination of $J^P$. We therefore keep the two model predictions, $5/2^+$ and $5/2^-$, without selecting a unique assignment.

In the case of  the excited $\Omega$ baryons, a constituent quark model study  assigns $\Omega(2250)$, $\Omega(2380)$, and $\Omega(2470)$ as $1D~(5/2^+)$, $2P~(1/2^-)$, and $3S~(3/2^+)$ states, respectively \cite{Menapara:2021vug}. An independent chiral quark model analysis is also compatible with $1D$, $J^P=5/2^+$ interpretation for $\Omega(2250)$ \cite{Xiao:2018pwe}. Our model predicts

\begin{equation*}
J^P(\Omega(2250))=\frac{3}{2}^{+},\qquad
J^P(\Omega(2380))=\frac{3}{2}^{-},\qquad
J^P(\Omega(2470))=\frac{3}{2}^{+}.
\end{equation*}

Therefore, the $\Omega(2470)$ prediction agrees exactly with the first study, while the predictions for $\Omega(2250)$ and $\Omega(2380)$ do not agree.

Concerning $\Omega(2012)^-$, both a chiral quark model analysis \cite{Xiao:2018pwe} and a recent Hamiltonian effective field theory analysis \cite{Han:2025gkp} predict the state to have $J^P=3/2^-$. Moreover, QCD sum rules calculation also support negative parity, but they cannot clearly distinguish between $1/2^-$ and $3/2^-$ \cite{Su:2024lzy}. Our prediction $ J^P=\frac{3}{2}^{-} $
is therefore consistent with several theoretical models. For $\Omega(2109)$, the situation is more model dependent: a recent quark model analysis suggests $J^P=3/2^+$ \cite{Luo:2025cqs}, whereas a coupled channel study favors $J^P=1/2^-$ \cite{Veloso:2026kii}. Our negative parity prediction  $J^P=1/2^-$  agrees with the coupled channel study, while our positive parity  prediction,$J^P=1/2^+$, differs from the quark model assignment. Therefore, agreement is found only for the negative parity branch.

Regarding $\Lambda(2585)$, the spin-parity quantum numbers remain poorly constrained. Our model predicts $J=5/2$ for positive and negative parity. Since no sufficiently reliable theoretical or experimental assignment has been identified to distinguish between these two possibilities, we do not select a unique $J^P$ prediction for this state.

\begin{table}[h!]

\centering

\begin{tabular}{l c c}

\hline

State & Our model & Theory \\

\hline

$\Xi(1620)$ & $\frac{1}{2}^-$
& $\frac{1}{2}^-$ \cite{Ramos:2002xh,Feijoo:2023wua} \\ \hline

$\Xi(1690)$ & $\frac{1}{2}^-$
& $\frac{1}{2}^-$ \cite{Sekihara:2015qqa,Feijoo:2023wua} \\ \hline

$\Xi(2120)$ & $\frac{3}{2}^{\pm}$
& $\frac{1}{2}^-,\,\frac{3}{2}^-,\,\frac{3}{2}^+$
\cite{Huang:2018uox,Khemchandani:2017wfb,Hei:2023eqz} \\ \hline

$\Xi(1950)$ & $\frac{1}{2}^-$
& $\frac{1}{2}^-,\,\frac{5}{2}^+,\,\frac{5}{2}^-$
\cite{PavonValderrama:2011gp,Xiao:2013xi,Yan:2024usf} \\ \hline

$\Sigma(2250)$ & $\frac{5}{2}^-$
& $\frac{3}{2}^-$ \cite{Ben:2025vfw} \\ \hline

$\Sigma(2455)$ & $\frac{1}{2}^-$
& $\frac{1}{2}^-$ \cite{Ben:2025vfw} \\ \hline

$\Sigma(2620)$ & $\frac{1}{2}^-$
& $\frac{1}{2}^-$ \cite{Ben:2025vfw} \\ \hline

$\Xi(2250)$ & $\frac{3}{2}^+$
& $\frac{3}{2}^+$ \cite{Ben:2025vfw} \\ \hline

$\Xi(2370)$ & $\frac{3}{2}^-$
& $\frac{3}{2}^-$ \cite{Ben:2025vfw} \\ \hline

$\Xi(2500)$ & $\frac{3}{2}^\pm$
&  $J^P$ unknown\\ \hline

$\Sigma(3000)$ & $\frac{1}{2}^-$
& $\frac{5}{2}^-$ \cite{Kim:2026ejb} \\ \hline

$\Sigma(3170)$ & $\frac{5}{2}^{\pm}$
& $J^P$ unknown \\ \hline

$\Omega(2250)$ & $\frac{3}{2}^+$
& $\frac{5}{2}^+$ \cite{Menapara:2021vug,Xiao:2018pwe} \\ \hline

$\Omega(2380)$ & $\frac{3}{2}^-$
& $\frac{1}{2}^-$ \cite{Menapara:2021vug} \\ \hline

$\Omega(2470)$ & $\frac{3}{2}^+$
& $\frac{3}{2}^+$ \cite{Menapara:2021vug} \\ \hline

$\Omega(2012)^-$ & $\frac{3}{2}^-$
& $\frac{3}{2}^-$ \cite{Xiao:2018pwe,Han:2025gkp};
$\frac{1}{2}^-,\,\frac{3}{2}^-$ \cite{Su:2024lzy} \\ \hline

$\Omega(2109)$ & $\frac{1}{2}^\pm$
& $\frac{3}{2}^+$ \cite{Luo:2025cqs};
$\frac{1}{2}^-$ \cite{Veloso:2026kii} \\ \hline

$\Lambda(2585)$
& $\frac{5}{2}^\pm$
& $J^P$ unknown \\ \hline

\hline

\end{tabular}

\caption{Comparison of the model predictions with representative theoretical
assignments for light strange baryons.}

\label{tab:strange_baryon_comparison}

\end{table}

\subsection{Heavy baryons and hidden charm pentaquarks}

Now, we proceed to present our results for heavy baryons and hidden charm pentaquarks and  compare  the model predictions with representative theoretical
assignments (see Table \ref{tab:heavy_baryon_comparison}). For $\Xi_c(3080)$, several theoretical studies predicted the   positive parity $D$-wave interpretation with $J^P=5/2^+$. This assignment is supported by constituent quark model calculations \cite{Yao:2018jmc}, QCD sum rules \cite{Wang:2017vtv}, and a relativistic flux tube analysis \cite{Chen:2014nyo}. However, an earlier chiral quark model calculation predicted a radially excited $2S$ state with $J^P=1/2^+$ \cite{Liu:2012sj}, thus, the identification is not unique. In our model, we obtain $J^P=3/2^+$ 
 which dose not agree with  either $J=5/2$ or $J=1/2$ as proposed in the cited theoretical studies.

Regarding $\Xi_b(6087)^0$ and $\Xi_b(6095)^0$, QCD sum rule calculations assign the two states to the $\Lambda$ type $P$-wave bottom baryon multiplet \cite{Xin:2023usd} with 

\begin{equation*}
J^P(\Xi_b(6087)^0)=\frac{1}{2}^{-},\qquad
J^P(\Xi_b(6095)^0)=\frac{3}{2}^{-}.
\end{equation*}

 An independent HQET based QCD sum rule analysis had also predicted a $J^P=1/2^-$ partner near $6.08$ GeV for the $J^P=3/2^-$ $\Xi_b(6100)^-$ state, which was identified later with the observed $\Xi_b(6087)^0$ \cite{Yang:2022oog,Tan:2023opd}. Our model predicts
\begin{equation*}
J^P(\Xi_b(6087)^0)=\frac{1}{2}^{-},\qquad
J^P(\Xi_b(6095)^0)=\frac{3}{2}^{-},
\end{equation*}
in exact agreement with these theoretical assignments.

The $\Xi_b(6227)$ is studied within  several theoretical approaches. For example, QCD sum rule calculations predict the $P$-wave bottom baryon  with $J^P=3/2^-$ \cite{Cui:2019dzj}. Moreover, a constituent quark model admits $J^P=3/2^-$ or $5/2^-$ \cite{Wang:2018fjm}. Molecular descriptions have also proposed a $J^P=1/2^-$ assignment \cite{Huang:2018bed}, indicating that its quantum numbers are not yet uniquely established. A dedicated QCD sum rule study of $\Xi_b(6227)^0$, which is interpreted as an  isospin partner of $\Xi_b(6227)^-$, supports a $1P$, $J=3/2$ interpretation \cite{Azizi:2020azq}.
For $\Xi_b(6227)$, the negative parity branch of our model gives $J^P=3/2^-$. 
This result is consistent with the assignments presented in QCD sum rules studies  \cite{Azizi:2020azq,Cui:2019dzj} as well as with one from constituent quark models \cite{Wang:2018fjm}. Although, it differs from the $1/2^-$ assignment which is considered within the molecular interpretation in Ref.~\cite{Huang:2018bed} .

The two narrow states $\Xi_b(6327)^0$ and $\Xi_b(6333)^0$ are widely interpreted as members of a $1D$ $\Xi_b$ doublet. A chiral quark model analysis assigns \cite{Wang:2022zcy}
\begin{equation*}
J^P(\Xi_b(6327)^0)=\frac{3}{2}^{+},\qquad
J^P(\Xi_b(6333)^0)=\frac{5}{2}^{+}.
\end{equation*}

 The same assignments are also confirmed  by QCD sum rule calculations \cite{Yu:2021zvl} and by a later quark pair creation analysis \cite{Zhou:2023xxs}. Our model predicts $J^P=3/2^+$ for both states. Consequently, the prediction for $\Xi_b(6327)^0$ is in exact agreement with these theoretical assignments, while for $\Xi_b(6333)^0$  the predicted spin is $3/2$ rather than $5/2$.

The $\Sigma_c(2800)$ is considered to be  a negative parity $P$-wave state, although its spin remains model dependent. QCD sum rule calculations suggest the existence of  $J^P=1/2^-$ and $3/2^-$ states \cite{Chen:2015kpa}, while constituent quark model calculations indicate $3/2^-$ or $5/2^-$ as possibilities \cite{Wang:2017kfr}. A $J^P=1/2^-$ molecular interpretation has also been considered \cite{Zhang:2012jk}. Our negative parity branch gives $ J^P=\frac{3}{2}^{-} $
that is consistent with some of the theoretical interpretations, but the spin-parity assignment is not uniquely established.

Regarding $\Sigma_b(6097)$, several theoretical studies predicted a negative parity $P$-wave assignment with $J^P=3/2^-$. Light cone QCD sum rule analyses support this picture \cite{Aliev:2018vye,Cui:2019dzj}, while the constituent quark model and $^3P_0$ decay studies predict $J^P=3/2^-$ or $5/2^-$ \cite{Wang:2018fjm,Yang:2018lzg}. Our negative parity branch predicts
$ J^P=\frac{3}{2}^{-} $
Thus, the model is in agreement with the theoretical expectation.

The hidden charm pentaquarks provide useful comparisons with several molecular and QCD based calculations. The $P_c(4312)$ is commonly identified as an $S$-wave $\Sigma_c\bar D$ molecule with $J^P=1/2^-$ \cite{Xiao:2019mvs,Chen:2019bip,Wang:2019got}. Our model predicts $ J^P=\frac{1}{2}^{-} $
consistent with this widely accepted scenario.
For $P_c(4440)$ and $P_c(4457)$, the spin ordering remains model dependent. One molecular scenario assigns $J^P=1/2^-$ and $3/2^-$ to $P_c(4440)$ and $P_c(4457)$, respectively \cite{Xiao:2019mvs}, whereas heavy quark spin and QCD sum rule analyses allow alternative assignments, including $J^P=3/2^-$ for both states \cite{Wang:2019got,Shimizu:2019ptd}. Our model predicts
\begin{equation*}
J^P(P_c(4440))=\frac{3}{2}^{-},\qquad
J^P(P_c(4457))=\frac{3}{2}^{-}.
\end{equation*}

Therefore, the $P_c(4457)$ prediction is consistent with the frequently used $3/2^-$ assignment, whereas the $P_c(4440)$ result is compatible with some theoretical scenarios but not with those that suggest $1/2^-$.
As to the $P_c(4380)$ state, several molecular and QCD sum rule studies suggest the $\Sigma_c^*\bar D$ configuration with $J^P=3/2^-$ \cite{He:2015cea, Wang:2018waa}. Our model  gives $ J^P(P_c(4380))=\frac{3}{2}^{-} $
and thus  the resulting value of the spin  agrees with these interpretations.
For the $P_{cs}(4459)^0$ state, both $J^P=1/2^-$ and $3/2^-$ molecular configurations have been studied in \cite{Peng:2020hql,Xiao:2021rgp}. However, coupled channel and one boson exchange approaches specifically predict a $J^P=3/2^-$ $\Xi_c\bar D^*$-dominated state \cite{Chen:2021tip}, while other analyses also find phenomenological support for the $J=3/2$ solution \cite{Du:2021bgb}. Our prediction $J^P(P_{cs}(4459)^0)=\frac{3}{2}^{-}  $ is  in good agreement with several independent molecular interpretations, although the alternative $1/2^-$ assignment has not been excluded.

\begin{table}[h!]

\centering

\begin{tabular}{l c c}

\hline

State & Our model & Theory \\

\hline

$\Xi_c(3080)$ & $\frac{3}{2}^+$
& $\frac{5}{2}^+$ \cite{Yao:2018jmc,Wang:2017vtv,Chen:2014nyo};
$\frac{1}{2}^+$ \cite{Liu:2012sj} \\ \hline

$\Xi_b(6087)^0$ & $\frac{1}{2}^-$
& $\frac{1}{2}^-$ \cite{Xin:2023usd,Yang:2022oog,Tan:2023opd} \\ \hline

$\Xi_b(6095)^0$ & $\frac{3}{2}^-$
& $\frac{3}{2}^-$ \cite{Xin:2023usd} \\ \hline

$\Xi_b(6227)$ & $\frac{3}{2}^-$
&$ \frac{1}{2}^- $ \cite{Huang:2018bed},
 $\frac{3}{2}^-$ \cite{Cui:2019dzj,Azizi:2020azq,Wang:2018fjm},
$\frac{5}{2}^-$ \cite{Wang:2018fjm} \\ \hline

$\Xi_b(6327)^0$ & $\frac{3}{2}^+$
& $\frac{3}{2}^+$ \cite{Wang:2022zcy,Yu:2021zvl,Zhou:2023xxs} \\ \hline

$\Xi_b(6333)^0$ & $\frac{3}{2}^+$
& $\frac{5}{2}^+$ \cite{Wang:2022zcy,Yu:2021zvl,Zhou:2023xxs} \\ \hline

$\Sigma_c(2800)$ & $\frac{3}{2}^-$
& $\frac{1}{2}^-,\,\frac{3}{2}^-$ \cite{Chen:2015kpa},
$\frac{3}{2}^-,\,\frac{5}{2}^-$ \cite{Wang:2017kfr},
$\frac{1}{2}^-$ \cite{Zhang:2012jk} \\ \hline

$\Sigma_b(6097)$ & $\frac{3}{2}^-$
& $\frac{3}{2}^-$ \cite{Aliev:2018vye,Cui:2019dzj},
$\frac{3}{2}^-,\,\frac{5}{2}^-$ \cite{Wang:2018fjm,Yang:2018lzg} \\ \hline

$P_c(4312)$ & $\frac{1}{2}^-$
& $\frac{1}{2}^-$ \cite{Xiao:2019mvs,Chen:2019bip,Wang:2019got} \\ \hline

$P_c(4440)$ & $\frac{3}{2}^-$
& $\frac{1}{2}^-$ \cite{Xiao:2019mvs},
$\frac{3}{2}^-$ \cite{Shimizu:2019ptd,Wang:2019got} \\ \hline

$P_c(4457)$ & $\frac{3}{2}^-$
& $\frac{3}{2}^-$ \cite{Xiao:2019mvs},
$\frac{3}{2}^-$ \cite{Shimizu:2019ptd,Wang:2019got} \\ \hline

$P_c(4380)$ & $\frac{3}{2}^-$
& $\frac{3}{2}^-$ \cite{He:2015cea,Wang:2018waa} \\ \hline

$P_{cs}(4459)^0$ & $\frac{3}{2}^-$
& $\frac{1}{2}^-,\,\frac{3}{2}^-$
\cite{Peng:2020hql,Xiao:2021rgp},
$\frac{3}{2}^-$ \cite{Chen:2021tip,Du:2021bgb} \\ \hline

\hline

\end{tabular}

\caption{Comparison of the model predictions with representative theoretical
assignments for heavy baryons and hidden-charm pentaquark candidates.}

\label{tab:heavy_baryon_comparison}

\end{table}
\section {Summary and conclusion}\label{SC}

We have constructed a data-driven approach to predict the total angular momentum of hadronic states specifically those for which spin assignments ​​have not yet been experimentally established or for whose spin assignments remain poorly constrained. To develop this model, we utilized meson  and baryon states from the Particle Data Group (PDG) database, incorporating information such as quark composition, isospin, parity, mass, and other spectroscopic features. In addition to providing these predictions, we have compiled a systematic database of hadronic states whose total angular momentum ($J$) values ​​are either unknown or lack precise constraints. This database offers a defined set of states for future phenomenological, experimental, and data-driven research. Since parity is included as an input in our method, the model predicts $J$ under a given parity assumption.

Comparing our results with existing theoretical literature allows us to assess the extent to which our predictions align with assignment proposed by theoretical models. In this analysis, among 45 resonances considered in this study, 42  states had theoretical spin-parity values ​​available in the literature used for comparison. Our predictions for 36 of these 42 resonances (approximately  $ 86\% $) are consistent with at least one of the reported theoretical values; for 6 resonances, our predictions differ, while for 3 resonances, no theoretical values ​​are available. This correspondence encompasses various hadronic sectors and diverse theoretical approaches, including quark models,  QCD sum rules, molecular models, and others. It is worth noting that any discrepancies should not necessarily be interpreted as a failure of our model, as the quantum numbers for these states have not yet been experimentally determined. However, they may also reflect limitations of the model or uncertainties in the theoretical assignment. Furthermore, in some instances, there are even conflicting theoretical assignments for  their quantum numbers. Such states are of particular significance, as they represent alternative scenarios that can be further investigated in light of future experimental results. The findings indicate that the proposed method offers more than just isolated spin predictions; specifically, by providing a list of hadronic states with unresolved spin assignments along with predictions for the total angular momentum of each it establishes a foundation for further research in the field of hadron spectroscopy. In particular, the prediction of $J$ allows us to compare our results with spin assignments derived from other theoretical frameworks, such as quark models, QCD sum rules, molecular models, lattice QCD, and various others. Consequently, such a comparison enables the identification of hadronic states for which independent approaches yield consistent results, as well as those that warrant further investigation. Furthermore, this database provides a systematic list of unassigned hadronic states, which can be expanded as new experimental and theoretical information becomes available.

\end{document}